\documentstyle[twoside,psfig,amsmath]{article}

\catcode`\@=11
\long\def\@makefntext#1{
\protect\noindent \hbox to 3.2pt {\hskip-.9pt  
$^{{\eightrm\@thefnmark}}$\hfil}#1\hfill}		

\def\@makefnmark{\hbox to 0pt{$^{\@thefnmark}$\hss}}	
	
\def\ps@myheadings{\let\@mkboth\@gobbletwo
\def\@oddhead{\hbox{}
\rightmark\hfil\eightrm\thepage}   
\def\@oddfoot{}\def\@evenhead{\eightrm\thepage\hfil
\leftmark\hbox{}}\def\@evenfoot{}
\def\sectionmark##1{}\def\subsectionmark##1{}}

\oddsidemargin=\evensidemargin
\newcounter{sectionc}\newcounter{subsectionc}\newcounter{subsubsectionc}
\renewcommand{\section}[1] {\vspace{12pt}\addtocounter{sectionc}{1} 
\setcounter{subsectionc}{0}\setcounter{subsubsectionc}{0}\noindent 
	{\tenbf\thesectionc. #1}\par\vspace{5pt}}
\renewcommand{\subsection}[1] {\vspace{12pt}\addtocounter{subsectionc}{1} 
	\setcounter{subsubsectionc}{0}\noindent 
	{\bf\thesectionc.\thesubsectionc. {\kern1pt \bfit #1}}\par\vspace{5pt}}
\renewcommand{\subsubsection}[1] {\vspace{12pt}\addtocounter{subsubsectionc}{1}
	\noindent{\tenrm\thesectionc.\thesubsectionc.\thesubsubsectionc.
	{\kern1pt \tenit #1}}\par\vspace{5pt}}
\newcommand{\nonumsection}[1] {\vspace{12pt}\noindent{\tenbf #1}
	\par\vspace{5pt}}

\newcounter{appendixc}
\newcounter{subappendixc}[appendixc]
\newcounter{subsubappendixc}[subappendixc]
\renewcommand{\thesubappendixc}{\Alph{appendixc}.\arabic{subappendixc}}
\renewcommand{\thesubsubappendixc}
	{\Alph{appendixc}.\arabic{subappendixc}.\arabic{subsubappendixc}}

\renewcommand{\appendix}[1] {\vspace{12pt}
        \refstepcounter{appendixc}
        \setcounter{figure}{0}
        \setcounter{table}{0}
        \setcounter{lemma}{0}
        \setcounter{theorem}{0}
        \setcounter{corollary}{0}
        \setcounter{definition}{0}
        \setcounter{equation}{0}
        \renewcommand{\thefigure}{\Alph{appendixc}.\arabic{figure}}
        \renewcommand{\thetable}{\Alph{appendixc}.\arabic{table}}
        \renewcommand{\theappendixc}{\Alph{appendixc}}
        \renewcommand{\thelemma}{\Alph{appendixc}.\arabic{lemma}}
        \renewcommand{\thetheorem}{\Alph{appendixc}.\arabic{theorem}}
        \renewcommand{\thedefinition}{\Alph{appendixc}.\arabic{definition}}
        \renewcommand{\thecorollary}{\Alph{appendixc}.\arabic{corollary}}
        \renewcommand{\theequation}{\Alph{appendixc}.\arabic{equation}}
        \noindent{\tenbf Appendix \theappendixc #1}\par\vspace{5pt}}
\newcommand{\subappendix}[1] {\vspace{12pt}
        \refstepcounter{subappendixc}
        \noindent{\bf Appendix \thesubappendixc. {\kern1pt \bfit #1}}
	\par\vspace{5pt}}
\newcommand{\subsubappendix}[1] {\vspace{12pt}
        \refstepcounter{subsubappendixc}
        \noindent{\rm Appendix \thesubsubappendixc. {\kern1pt \tenit #1}}
	\par\vspace{5pt}}

\newcommand{\textlineskip}{\baselineskip=13pt}
\newcommand{\smalllineskip}{\baselineskip=10pt}

\def\eightcirc{
\begin{picture}(0,0)
\put(4.4,1.8){\circle{6.5}}
\end{picture}}
\def\eightcopyright{\eightcirc\kern2.7pt\hbox{\eightrm c}} 

\newcommand{\copyrightheading}[1]
	{\vspace*{-2.5cm}\smalllineskip{\flushleft
	{\footnotesize International Journal of Modern Physics B, #1}\\
	{\footnotesize $\eightcopyright$\, World Scientific Publishing
	 Company}\\
	 }}

\newcommand{\publisher}[2]{{\begin{center}\footnotesize\smalllineskip 
	Received #1\\
	Revised #2
	\end{center}
	}}

\def\abstracts#1#2#3{{
	\centering{\begin{minipage}{4.5in}\baselineskip=10pt\footnotesize
	\parindent=0pt #1\par 
	\parindent=15pt #2\par
	\parindent=15pt #3
	\end{minipage}}\par}} 

\renewenvironment{thebibliography}[1]			
	{\frenchspacing
	 \ninerm\baselineskip=11pt
	 \begin{list}{\arabic{enumi}.}
	{\usecounter{enumi}\setlength{\parsep}{0pt}
	 \setlength{\leftmargin 12.7pt}{\rightmargin 0pt} 
	 \setlength{\itemsep}{0pt} \settowidth
	{\labelwidth}{#1.}\sloppy}}{\end{list}}

\newcounter{itemlistc}
\newcounter{romanlistc}
\newcounter{alphlistc}
\newcounter{arabiclistc}

\newcommand{\fcaption}[1]{
        \refstepcounter{figure}
        \setbox\@tempboxa = \hbox{\footnotesize Fig.~\thefigure. #1}
        \ifdim \wd\@tempboxa > 5in
           {\begin{center}
        \parbox{5in}{\footnotesize\smalllineskip Fig.~\thefigure. #1}
            \end{center}}
        \else
             {\begin{center}
             {\footnotesize Fig.~\thefigure. #1}
              \end{center}}
        \fi}

\newcommand{\tcaption}[1]{
        \refstepcounter{table}
        \setbox\@tempboxa = \hbox{\footnotesize Table~\thetable. #1}
        \ifdim \wd\@tempboxa > 5in
           {\begin{center}
        \parbox{5in}{\footnotesize\smalllineskip Table~\thetable. #1}
            \end{center}}
        \else
             {\begin{center}
             {\footnotesize Table~\thetable. #1}
              \end{center}}
        \fi}

\def\@citex[#1]#2{\if@filesw\immediate\write\@auxout
	{\string\citation{#2}}\fi
\def\@citea{}\@cite{\@for\@citeb:=#2\do
	{\@citea\def\@citea{,}\@ifundefined
	{b@\@citeb}{{\bf ?}\@warning
	{Citation `\@citeb' on page \thepage \space undefined}}
	{\csname b@\@citeb\endcsname}}}{#1}}

\newif\if@cghi
\def\cite{\@cghitrue\@ifnextchar [{\@tempswatrue
	\@citex}{\@tempswafalse\@citex[]}}
\def\citelow{\@cghifalse\@ifnextchar [{\@tempswatrue
	\@citex}{\@tempswafalse\@citex[]}}
\def\@cite#1#2{{$\null^{#1}$\if@tempswa\typeout
	{IJCGA warning: optional citation argument 
	ignored: `#2'} \fi}}

\def\pmb#1{\setbox0=\hbox{#1}
	\kern-.025em\copy0\kern-\wd0
	\kern.05em\copy0\kern-\wd0
	\kern-.025em\raise.0433em\box0}

\def\fnt#1#2{\footnotetext{\kern-.3em
	{$^{\mbox{\scriptsize #1}}$}{#2}}}

\def\fpage#1{\begingroup
\voffset=.3in
\thispagestyle{empty}\begin{table}[b]\centerline{\footnotesize #1}
	\end{table}\endgroup}

\def\runninghead#1#2{\pagestyle{myheadings}
\markboth{{\protect\footnotesize\it{\quad #1}}\hfill}
{\hfill{\protect\footnotesize\it{#2\quad}}}}
\font\tenrm=cmr10
\font\tenit=cmti10 
\font\tenbf=cmbx10
\font\bfit=cmbxti10 at 10pt
\font\ninerm=cmr9

\font\eightrm=cmr8

\def\qed{\hbox{${\vcenter{\vbox{			
   \hrule height 0.4pt\hbox{\vrule width 0.4pt height 6pt
   \kern5pt\vrule width 0.4pt}\hrule height 0.4pt}}}$}}

\def\bsc{{\sc a\kern-6.4pt\sc a\kern-6.4pt\sc a}}	
\def\bflatex{\bf L\kern-.30em\raise.3ex\hbox{\bsc}\kern-.14em 
T\kern-.1667em\lower.7ex\hbox{E}\kern-.125em X} 

\newcommand{\gsim}{\raisebox{-1mm}{$\stackrel{\textstyle >}{\textstyle \sim}$}}

\begin{document}

\runninghead{G.~Falci and A.~Fubini}{Quantum fluctuations in superconducting dots 
at finite temperatures} 

\normalsize\textlineskip
\thispagestyle{empty}
\setcounter{page}{1}

\copyrightheading{}			

\vspace*{0.88truein}

\fpage{1}
\centerline{\bf QUANTUM FLUCTUATIONS IN SUPERCONDUCTING DOTS} 
\centerline{\bf AT FINITE
TEMPERATURES}
\vspace*{0.37truein}
\centerline{\footnotesize G. FALCI}
\vspace*{0.015truein}
\centerline{\footnotesize\it Dipartimento di Metodologie Fisiche e Chimiche 
(DMFCI), Universit\`a di Catania} 
\baselineskip=10pt
\centerline{\footnotesize\it   Viale A. Doria 6, Catania. 95125, Italy}

\centerline{\footnotesize\it Istituto Nazionale per la Fisica della Materia, 
Unit\`a di Catania (Italy)} 
\centerline{\footnotesize\it LEPES-CNRS, Grenoble (France)}
\baselineskip=10pt

\vspace*{10pt}
\centerline{\normalsize and}
\vspace*{10pt}
\centerline{\footnotesize A. FUBINI}
\vspace*{0.015truein}
\centerline{\footnotesize\it Dipartimento di Fisica dell'Universit\'a
degli Studi di Firenze} 
\baselineskip=10pt
\centerline{\footnotesize\it and Istituto Nazionale per la Fisica
della Materia, Unit\`a di Firenze, }
\baselineskip=10pt
\centerline{\footnotesize\it Largo E.~Fermi~2, 50125 Firenze, Italy }
\vspace*{0.225truein}
\publisher{(received date)}{(revised date)}

\vspace*{0.21truein}
\abstracts{
We study the thermodynamics of ultrasmall metallic grains with level
spacing $\delta$ comparable or smaller than the pairing correlation
energy, at finite temperatures, $T \gsim \delta$. We describe a method 
which allows to find quantum corrections
to the effect of classical fluctuations. We present results for thermodynamic
quantities in ordered grains and for the reentrant odd susceptibility 
in disordered grains.}{}{}

\section{Outline of the problem}
Superconducting coherence in finite systems has recently received 
a renewed attention since the series of 
experiments of Ralph Black and Tinkham (RBT).\cite{RBT} They succeeded
to perform tunneling spectroscopy on {\em individual} nanoscale $Al$ grains 
where  $E_C$, the charging energy\cite{kn:Averin-Likharev-91}, is very
large (up to $500 \, K$), enforcing the conservation of 
the total number of electrons $N$. They observed 
spectroscopic features related to superconductivity, which also depended
on the parity ($N$ being odd or even) of the grains. 

The question of the size limit for a metal particle to have superconducting
properties was posed since the early days of BCS theory by
Anderson.\cite{Anderson59} It was argued that when the average level spacing 
$\delta \sim 1/({\cal N}(0)V)$ becomes of the order of the BCS gap $\Delta$, 
superconductivity should disappear. 

In small metallic grains\cite{Perenboom81} thermal fluctuations 
wash out the hallmarks of Cooper pair condensation,
like the zero resistance and the complete Meissner effect.
However other  
features related to superconductivity can be clearly observed.
In micrometer size samples charging effects determine the Coulomb
Blockade\cite{kn:Averin-Likharev-91} and parity effects\cite{parity,parity1}, 
but superconducting correlations are still strong 
(the typical energy  scale is the BCS $\Delta$).
For ultrasmall grains ($\delta \geq \Delta$, ``superconducting dots'') 
at low temperature, quantum fluctuations due to both large $E_C$ and 
finite $\delta$ tend to suppress superconducting properties. 

Theoretical analysis by parity dependent mean field theories\cite{parity1}
found  that the breakdown of BCS superconductivity occurs at a value of 
$\delta / \Delta$ which depends on parity.
The use of the mean field grand-canonical approach is appropriate
when the grains are large enough to give small relative
fluctuations of the electron number (this condition is not strictly
met in the experiments\cite{RBT}) and it breaks down completely
in superconducting dots at $T=0$. In the canonical ensemble the BCS 
pair amplitude is zero and in order to characterize pairing correlations 
different quantities have to be studied\cite{vonDelft}, 
for instance the odd-even staggering
between the ground state energies at different 
$N$.\cite{Matveev97,Mastellone98,Berger98} They studied the BCS 
Hamiltonian
\begin{eqnarray}
        {\cal H} &=& \sum_{{n, \sigma}}
          (\epsilon_n - \sigma \mu_B H)\,
        c_{n,\sigma}^{\dagger}  c_{n,\sigma}  
\;-\; \lambda
        \sum_{m,n=1}^{\Omega}c_{m,+}^{\dagger}c_{m,-}^{\dagger}
        c_{n,-}c_{n,+} \hskip10mm .
\label{hamiltonian}
\end{eqnarray}
where $\{m,n\}$ span a shell of $\Omega$ doubly degenerate time reversed
($\sigma=\pm$)
single particle energy levels
(energy $\epsilon_m$ and annihilation operator $c_{m,\sigma}$),
and $\lambda$ is the BCS coupling constant.
The effect of a magnetic field $H$ is accounted for by the 
Zeeman term ($\mu_B$ is the Bohr magneton). 

For equally spaced $\epsilon_m=m \delta$, it was found that even in 
ultrasmall grains pairing correlations
manifest with non-perturbative fluctuations\cite{Matveev97} and that 
the low-energy physics can be expressed by universal functions of 
$\delta/\Delta$ ($\Delta$ plays the role of an energy scale analogous 
to the Kondo temperature), allowing a quantitative description of the full
crossover regime.\cite{Mastellone98} 

The model Eq.(\ref{hamiltonian}) was further studied with various
methods\cite{kn:variuos}
until it was recognized that it has been exactly solved 
long ago by Richardson and Sherman.\cite{RICHARDSON} The connections
of this solution with conformal field theories and off-shell Bethe 
Ansatz\cite{SIERRA} has been recently exploited, and may yield
exact expressions for the correlation functions. 

It is in general very difficult to achieve experimentally unambiguous
evidence of the presence of strong pairing correlations in 
ultrasmall grains.
However the spin susceptibility in 
grains with $N$ odd $\chi_o(T)$ was found\cite{DiLorenzo99} 
to show a striking reentrant behavior 
(see Fig.1). This qualitative difference with normal metal grains
is a signature of pairing and persists even in ultrasmall grains. 
It is due to the interplay of 
pairing which tends to suppress
$\chi_o(T)$ for decreasing $T$, and parity which determines a Curie 
$1/T$ contribution which eventually prevails.

The result was obtained by combining a parity-projected path-integral 
approach in the static-path approximation (SPA)\cite{Muelschlegel} 
with low temperature results obtained by massive use of the exact solution. 
In this paper we describe
how to study quantum corrections to the SPA 
with a functional procedure which has been named RPA'.
The RPA' also allows to 
study the thermodynamics of 
weakly disordered grains at finite temperature, in an efficient and reliable way. 
This is an important virtue 
because experiments should be carried out in ensembles of grains, 
where a distribution of shape and volume has to be considered.

We remark that the Hamiltonian Eq.(\ref{hamiltonian}) describes
universal properties of small grains in the metallic regime (weak
disorder)\cite{kn:universal}, if we  
choose $\epsilon_m$ as the eigenvalue of a random matrix\cite{kn:metha}  
of the Gaussian Orthogonal Ensemble (GOE), since we 
ignore here spin-orbit effects.\cite{vonDelft}
A set of mono dispersed grains
should be studied along these lines. This makes practically impossible the
use of the exact solution as in the work of 
Di Lorenzo et al.\cite{DiLorenzo99}.
Some effects of disorder have been previously studied using a 
parity-projected mean field 
approach\cite{kn:smith-ambegaokar-96} valid for large grains, and 
at $T=0$.\cite{Sierra99} 

To clarify the virtues and differences of various approaches we 
sketch the procedure we employ. We start from a path-integral formulation
of the problem which we consider the grand-canonical ensembles
of odd or even grains.\cite{parity1} We express the partition function 
as a path integral over a quantum auxiliary field $\Delta(\tau)$. 
The SPA consists in retaining only the contribution of all constant 
$\Delta(\tau)=\Delta^{(s)}$, which is maximal if  
$\Delta^{(s)}=\Delta$, the BCS gap. Thus the SPA describes exactly 
classical fluctuations. We account for quantum fluctuations by allowing
for non constant paths  $\Delta(\tau)$. The RPA' consists in 
retaining Gaussian fluctuations along the static path by first 
fixing $\Delta^{(s)}$, then evaluating Gaussian fluctuations
and finally integrating over  
$\Delta^{(s)}$.\cite{kn:wang69,kn:keiter,kn:morandi74} More accurate 
approximation schemes are also possible. All these methods break down 
at low temperatures, in a more or less spectacular way. 
In this paper we use study the thermodynamics at finite temperature
beyond the SPA. We present results on the specific heat 
and on the odd spin susceptibility for equally spaced $\epsilon_n$. We  
finally check that the reentrant behavior of $\chi_o$ persists in  
disordered grains.

\section{Functional formulation}
The spin susceptibility and the specific heat of a grain with an even
(e) or an odd (o) number $N$ of electrons is defined starting from the
partition functions $Z(T,N)$, which should be evaluated in the
canonical ensemble. With the help of a parity projection
technique\cite{parity1} the grand partition function of an even or
odd system can be expressed as
\begin{align}
        Z_{e/o}(T,\mu) &= \frac12 \sum_{N=0}^{\infty} e^{ \mu N/T}
        [1 \pm e^{i \pi N}] Z(T,N) \notag \\
        &\equiv  \frac12 \left(Z_+ \pm Z_-\right) .
\label{e.Zeo}
\end{align}
Here, the partition function $Z_{+}$ is the usual grand partition function
at temperature $T$ and chemical potential $\mu_{+}=\mu$.  The grand
partition function $Z_{-}$ describes an auxiliary ensemble at
temperature $T$ and chemical potential $\mu_{-}=\mu + i\pi T$; it is a
formal tool, necessary to include parity effects. In the calculation,
the chemical potential $\mu$ will be placed between the topmost
occupied level and the lowest unoccupied level in the even case, while
it will be at the singly occupied level in the odd case.

The method we use to perform the calculation is based on the
functional formalism and the Hubbard-Stratonovic (HS) approach, where
the two-body interaction
$c_{m,+}^{\dagger}c_{m,-}^{\dagger}c_{n,-}c_{n,+}$ is reduced to a
one-body interaction with a time dependent external field
$\Delta(\tau)$, which has to be averaged over a Gaussian weight.  So
that the problem switches to the treatment of the interaction vertex
between the fermionic and the HS field. The functional integral over the
fermionic degrees of freedom is Gaussian and can be easily performed.
After this step the grand partition functions $Z_{\pm}$
reads\cite{Nambu} 
\begin{equation}
        Z_{\pm}= {\cal N}\!\!\int\!\!
        {\cal D}^2 \Delta
        \exp\bigg\{\!\int \limits_{0}^{\beta}
        {d\tau \Big[\sum_m \!\Big(
        {\rm Tr} \ln
\hat{G}_{m,\pm}^{-1}-\frac{\epsilon_m-\mu +\mu_B H}{T}\Big)\!-\!
        \frac{\mid \Delta \mid^2}{\lambda}
        \Big]}\!\bigg\}~,
\label{hubz}
\end{equation}
where, $\beta =1/T$ and ${\cal N}$ is the normalization factor of the
Gaussian ``measure'' introduced by the HS procedure. The matrix Green function 
$\hat{G}_{m,\pm}$ is given by
\begin{equation}        
	\hat{G}^{-1}_{m,\pm}=
	\begin{pmatrix}
	-\partial_\tau-\epsilon_m+\mu +\mu_B H & \Delta(\tau) \\
	\\
	\Delta^\dagger(\tau) & -\partial_\tau+\epsilon_m-\mu +\mu_B H
	\end{pmatrix} \delta(\tau-\tau^\prime)~.
\end{equation}
It is worth to stress that the $\{\pm\}$ ensembles are characterized by
different time-periodicity of the Green function: in the $\{-\}$ case
the adding term to the chemical potential $i\pi T$ has been re-absorbed by
the gauge transformation on the fermionic variables $c_{m,\sigma}(\tau)=\exp(i\pi T\tau)
d_{k,\sigma}$ changing, in this way, the periodic boundary condition. In
particular the fermionic (bosonic) Matsubara periodicity is associated
to the $Z_+$ ($Z_-$) partition function. 

In order to deal with the non-linear functional integral~(\ref{hubz}),
it is useful to extract the static term by the HS field,
i.e. $\Delta(\tau) = \Delta(\omega_0) +
\sum_{n\neq 0} \Delta(\omega_n)
\exp(-i\omega_n t)\equiv \Delta^{(s)} + \delta\Delta(\tau)$. The
SPA\cite{Muelschlegel} approximation consists in taking into account
the only static contribution $\Delta^{(s)}$, so that the functional
integral is therefore reduced to an ordinary integral over the class
of ``straight paths'' in the time interval $[0,\beta]$. It is worth to
notice that $\Delta^{(s)}$ is just the average point of the path (the
{\em centroid} in Feynman language) and for this reason some analogies
with the Feynman variational approach\cite{Feynman} can be found. As
in the variational case\cite{Giachetti,Kleinart}, the SPA can be
improved by taking into account small-amplitude (quadratic)
fluctuation around the centroid. This is the so-called RPA',
introduced by Wang {\em et al.}\cite{kn:wang69} and refined later by
Keiter {\em et al.}\cite{kn:keiter} in order to study the Anderson
model (for an exhaustive review on this approximation see also
Ref.\cite{kn:morandi74}). In particular it turns out that RPA'
corresponds to a ring-diagram summation equivalent to the standard
random phase approximation (RPA), except that only non-zero-frequency
transfers at the vertices are retained in the ring-diagrams, while all
zero-frequency transfers are treated
exactly.\cite{kn:keiter,kn:morandi74}

The logarithm term in Eq.~(\ref{hubz}) can be expanded in the
following way
\begin{equation}        
\ln\hat{G}_{m,\pm}^{-1} = \ln{\hat{G}_{m,\pm}^{(s)^{-1}}} +
\sum_{\ell=1}^{\infty} \frac{(-1)^{\ell-1}}{\ell}
\big(\hat{G}_{m,\pm}^{(s)} \hat{V}\big)^\ell~,
\label{e.series}
\end{equation}
where 
\begin{equation*}
\hat{V} =	
	\begin{pmatrix}
	0 & \delta\Delta(\tau) \\
	\delta\Delta^\dagger(\tau) & 0
	\end{pmatrix} 
\delta(\tau-\tau^\prime)~.
\end{equation*}
Performing the trace operation and paying attention to the fact that the zero
frequency are ruled out from the sums, it is easy to find the 
first order contribution vanish. So the static path $\Delta^{(s)}$ is a local
extreme of the integrand.
The leading correction to the SPA is
quadratic in the fluctuation around the centroid. In RPA'
the partition function reads
\begin{equation}
Z_\pm = {\cal N} \int_0^\infty d\Delta^{(s)} 
\exp\big[\varphi_\pm^{{}^{\rm SPA}}(\Delta^{(s)}) +
\delta^2\!\varphi_\pm(\Delta^{(s)})\big] ~,
\label{ZRPA'}
\end{equation}
where
\begin{equation}
\varphi_\pm^{{}^{\rm SPA}}(\Delta^{(s)})= - \frac{s}{\lambda\,T} + \sum_{m,\sigma} \left[\ln 2{
\cosh\frac{E_m + \sigma\mu_B H}{2T}\brace\sinh\frac{E_m + \sigma\mu_B H}{2T}}
  -  \frac{\xi_m}{2T}\right]~,
\label{SPA}
\end{equation}
with $E^2_{m} =\xi^2_m + s$ and $\xi_m= \epsilon_m -\mu$, is the
standard SPA contribution\cite{DiLorenzo99,Muelschlegel}. The
quantum fluctuation correction to the free energy is 
\begin{equation*}
\exp\big[\delta^2\!\varphi_\pm(\Delta^{(s)})\big] = \int {\cal D}^2[\delta\Delta] 
\exp\big\{-\delta^2\!S_\pm[\delta\Delta(\tau)]\big\}~,
\end{equation*}
where the funcional $\delta^2\!S_\pm[\delta\Delta]$ containins
products of two Green's functions. The path integral can be easily
evaluated after the standard sum procedure over (fermionic and
bosonic) Matsubara frequency has been performed. After some 
algebra the following expression is found
\begin{equation}
\delta^2\!\varphi_\pm(\Delta^{(s)}) = - \frac12 \sum_{\omega_n > 0}
\ln\Big[A_{\pm}^2(\omega_n) - B_{\pm}^2(\omega_n)\Big]~,
\end{equation}
where
\begin{alignat}{4}
A_{\pm}(\omega_n)\! &=
\sum_{m,\sigma}g_m(\omega_n)\tanh^{\pm1}\!\!\frac{E_m + \sigma\mu_B
H}{2T} +\frac1\lambda~, & ~~~~ g_m(\omega_n) &= -\frac{\xi^2_m +
E^2_m}{2E_m(\omega_m^2 + 4E_m^2)}~, \notag \\
\notag \\
B_{\pm}(\omega_n)&= \sum_{m,\sigma} f_m(\omega_n)\tanh^{\pm1}\!\!
\frac{E_m + \sigma\mu_B H}{2T}~, &~~f_m(\omega_n)&= \frac{s}{2E_m(\omega_m^2 + 4E_m^2)}~. 
\label{varie}
\end{alignat}

By means of expressions~(\ref{ZRPA'}-\ref{varie}) the thermodynamical
quantity of the model can be numerically evaluated with a CPU time
much shorter than what needed for the exact solution.
This makes possible the analysis of the effect of a statistical distribution
of non interacting levels. In the next
section we will show our results for even specific heat and for the
reentrant susceptibility in the ordered level spacing case and we
compare them with previous results both analytic and exact.\cite{DiLorenzo99} 
This comparison, besides, allows to test the reliability of the RPA'
in dealing with the quantum fluctuations in ultrasmall grains. 
Finally we will show results for disordered
superconducting dots.

\section{Results}
\begin{figure}[b!]
\label{fig:susceptibility}
\vspace*{13pt}
\centerline{\psfig{height=5cm,file=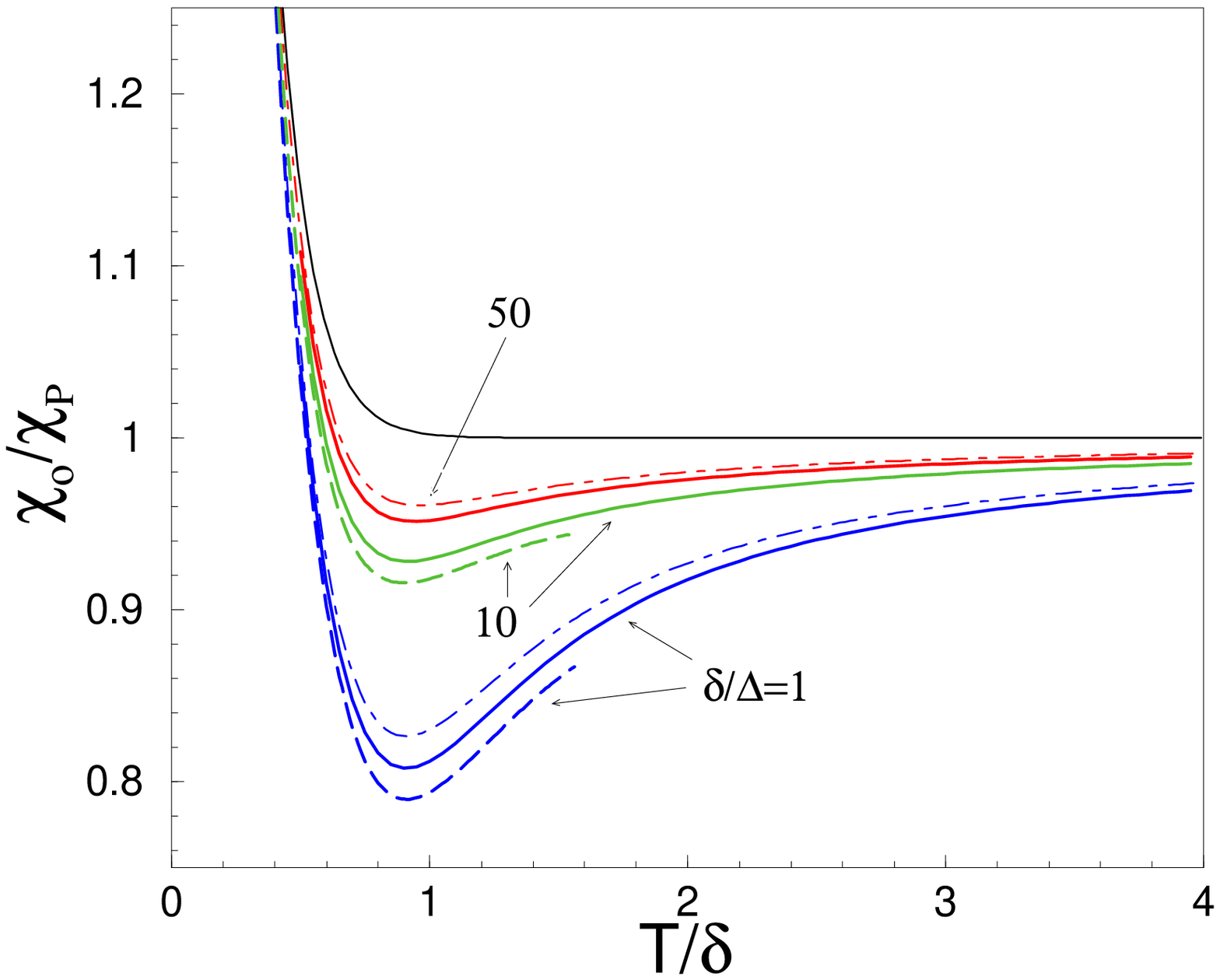}}
\fcaption{Spin susceptibility for an odd grain vs. temperature in
units of $\delta$. Solid line: RPA'; dashed line: exact; dot-dashed: SPA}
\end{figure}
Results obtained with RPA' are shown in Figs.1-3. 
We notice first of all that for $\chi_o(T)$ 
(Fig.1) the  RPA' curves (full lines) move away 
from the SPA curves 
(dot-dashed) towards the exact result (dashed curve). So the quantum 
corrections are appreciable and give the expected trend. The difference 
between RPA` and exact curves can be mainly attributed to differences 
between canonical ensemble and paruty-projected grand canonical 
ensemble.\cite{Muelschlegel2}  
The agreement at lower temperature is due to the fact that essentially 
only the classical Curie contribution is important. At very low temperature 
the SPA gives 
wrong results for the contribution of the ``condensate'' which is by itself 
small, whereas the RPA' breaks down. For even grains $\chi_e(T)$ vanishes
exponentially for $T \to 0$.

The specific heat for even grains is shown in Fig.2a. For large enough
coupling an anomaly develops. This is not the case for odd grains, as
expected, because the odd electron reduces the phase space for pairing
to be active. At lower temperatures the RPA' data seem to reach the
data obtained by diagonalization with a good quantitative accuracy, at
even lower temperatures RPA' breaks down. While is not surprising that
the Richardson and the parity projected RPA' curves do not match each
other at high temperature: the former arise from a correct canonical
analysis, the latter come from the parity projection technique, which
relies on the grand-canonical treatment. As shown by Denton {\em et
al.}\cite{Muelschlegel2} for a normal metal particle, the specific
heat in the grand-canonical case is lager than in the canonical case,
asymptotically is $\frac12 k_{\rm B}$ higher, because there are more
excitations allowed when the electron number is not conserved. In
order to point out this behavior, in Fig.2b $c_v \delta/\gamma T$ is
shown, where $\gamma = 2 \pi^2/3$ is the Sommerfeld coefficient of the
specific heat for an electron gas in the grand-canonical ensemble.
\begin{figure}[t!]
\label{fig:specific-heat}
\vspace*{13pt}
\psfig{height=5cm,file=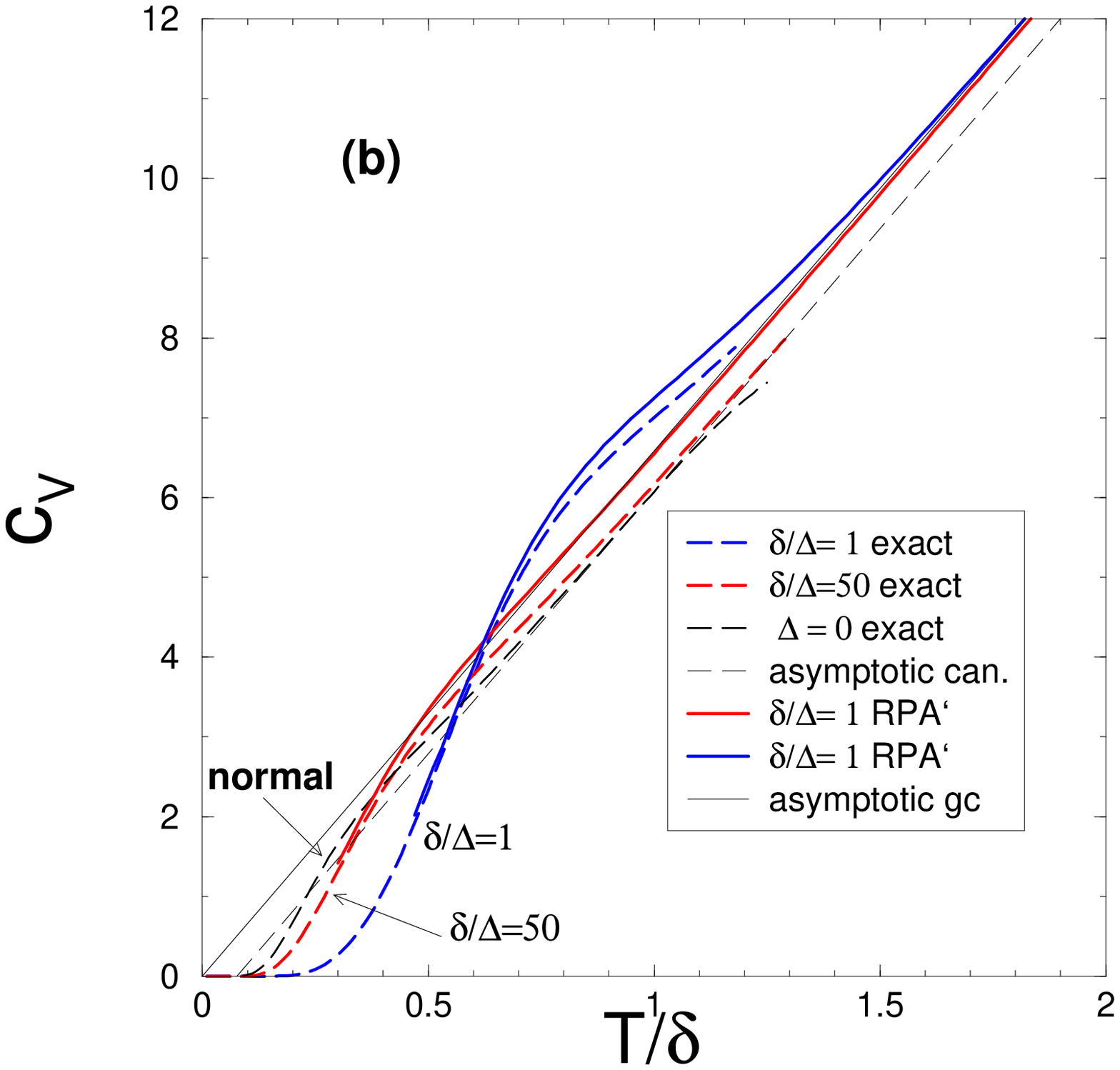}
\vspace*{-5cm}\hfill
\psfig{height=5cm,file=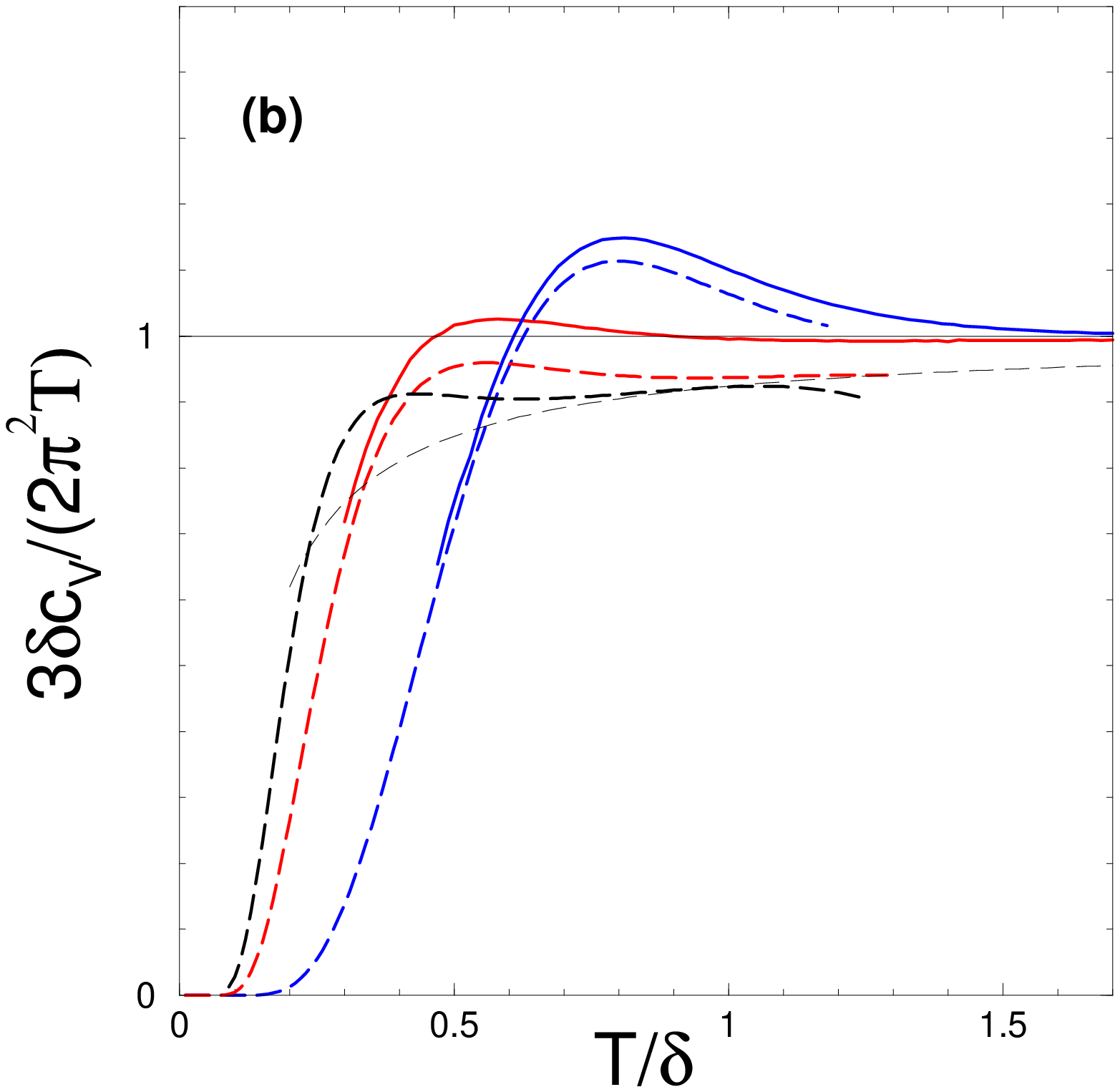}
\vspace*{13pt}
\fcaption{(a) Specific heath for an even grain vs. temperature in
units of $\delta$. (b) Rescaled specific heat $c_v \delta / \gamma T$ vs. temperature in
units of $\delta$ (see text.). Solid curves are the RPA' results whereas dashed
curves are the exact results. Thin lines represent the asymptotic behavior at
large $T$ in the canonical (dashed) and in the grandcanonocal (solid) 
ensemble.}
\end{figure}

\begin{figure}[h!]
\label{fig:chi-odd-disordered}
\vspace*{13pt}
\centerline{\psfig{height=5cm,file=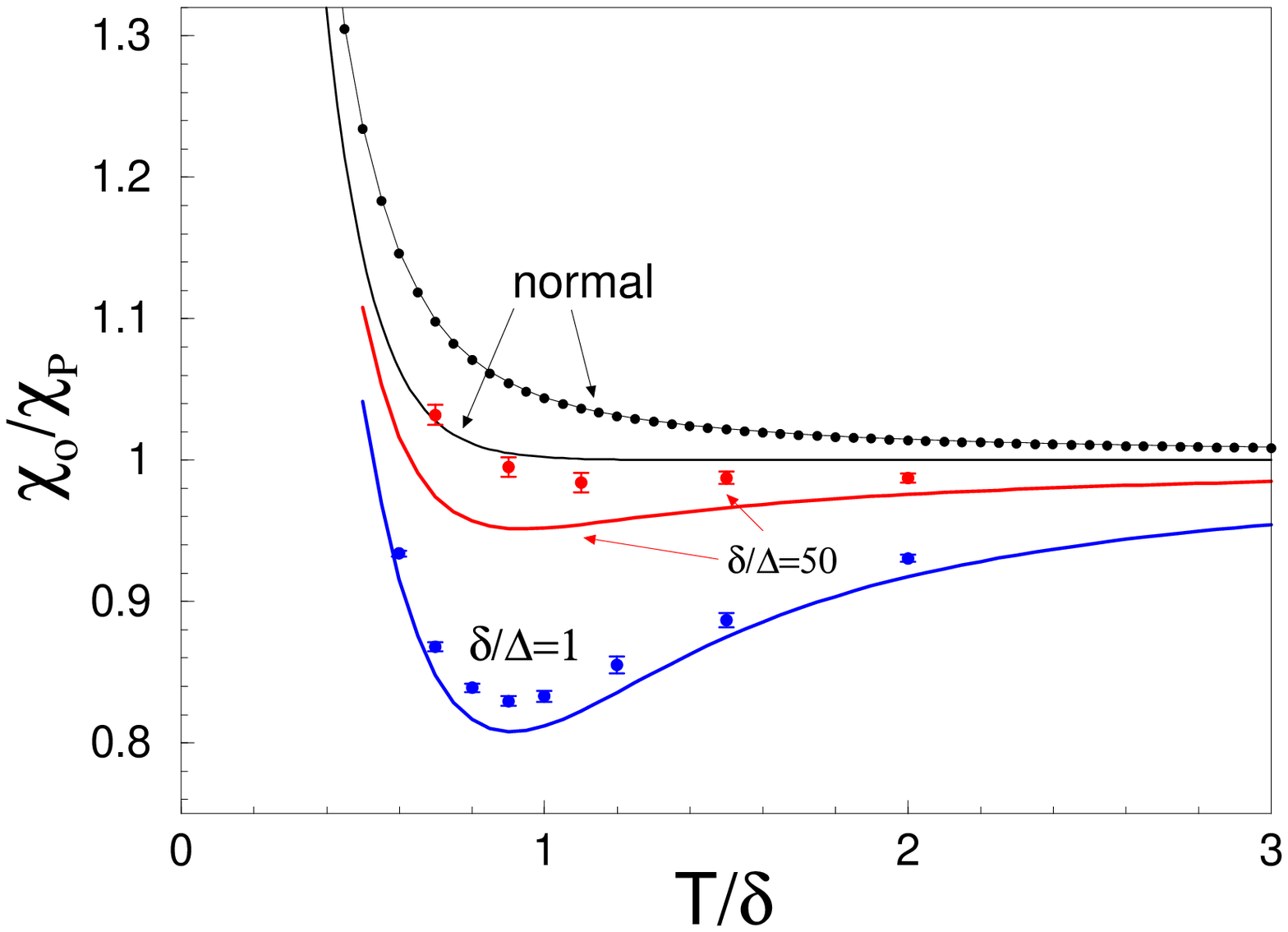}}
\vspace*{13pt}
\fcaption{Effects of a GOE statistical distribution of energy levels on
the reentrant behavior of $\chi_o(T)$. Solid lines represent the canonical
($\Delta=0$) and RPA` ($\delta/\Delta=1$) results for equally spaced 
noninteracting single-particle levels, while dots are the result for GOE 
distributed single-particle levels}
\end{figure}
Experimental investigation on thermodynamic properties of ultrasmall 
superconducting grains can only be performed in large ensembles 
rather than in individual grains. In the case of the susceptibility
it is of course not possible to prepare an ensemble of all odd grains.
However at temperatures where the Curie contribution prevails 
the susceptibility of even grains is already exponentially small.
A similar line of reasoning applies for the measurement of the even specific 
heat anomaly, which however seems to be less promising.
Two questions can be raised for what concerns the physics of ensembles of
grains. The first one concerns interaction between neighboring grains with 
possible electron tunneling. However for small enough grains large charging
energies, as the ones in the RBT\cite{RBT} experiment,
prevent charge transfer, except for very special situations. Another question
is the effect of the disorder. To discuss this issue we considered samples 
of monodispersed grains, i.e. grains differ in the shape but not in the volume.
This has striking consequences in small metallic grains, since 
shape irregularities over lengths $\sim \lambda_F$ are enough to 
affect the spectrum at the Fermi energy. Gorkov and Eliashberg\cite{Gork-Elias}
incorporated the effect of the disordered boundary conditions into the 
Hamiltonian, where they appear as random matrix elements which determine
a statistical distribution of the energy level. For vanishing magnetic field
the proper ensemble is the GOE\cite{kn:metha}. The thermodynamics of normal
disordered grains was studied by Denton {\em et al.}\cite{Muelschlegel2}

We applied the RPA' to this problem. Results are shown in Fig2., where
points represent ensemble averages over 200 samples of random
Hamiltonians belonging to the GOE. The spectra were obtained by
diagonalizing $500\times 500$ matrices with random elements and taking
the central 100 levels of the resulting semicircular distribution, in
order to prevent undesired border effects.  Results show that
$\chi_o(T)$ is clearly reentrant even in disordered samples.

\section{Conclusions}
We have discussed some thermodynamic properties of ultrasmall superconducting 
grains at $T>\delta$ using the RPA' approximation. 
Comparison with results in the SPA, which accounts exactly for classical 
fluctuations, allows to estimate the effect of quantum fluctuations. 
Comparison with the behavior of the exact solution results at $T\sim \delta$ 
allows to estimate differences between results in the grand canonical and in 
the canonical ensemble.

The RPA' turns out to be an efficient and reliable method. It can be 
successfully applied to problems for which the exact solution is not useful
in practice. An example is the study of the effect of a statistical 
distribution of noninteracting levels on the thermodynamics. We 
considered the odd susceptibility $\chi_o(T)$ which was found to show a 
characteristic reentrant behavior for grains with a regular spectrum. 
This is a promising quantity for experimental detections of paring 
correlations in ultrasmall grains. 
We checked that the reentrant behavior
is well preserved even in samples containing a large number of 
monodispersed grains,
whose spectrum has statistical properties described by the GOE.

\nonumsection{Acknowledgements}
\noindent
We acknowledge Rosario Fazio for several important discussions, 
A. Mastellone for providing the sets of GOE levels, A. Di Lorenzo
for technical remarks. We also acknowledge discussions with L. Amico,
G. Giaquinta, E. Piegari, G. Sierra, V. Tognetti. 
We acknowledge support of MURST (Cofinanziamento SCQBD) and 
INFM (PAIS-ELMAMES). GF acknowledges support
of EU (grant TMR-FMRX-CT-97-0143).

\nonumsection{References}

\end{document}